\documentstyle{twocolprepr}

\newcommand{\sigmahat}{\hat\sigma}
\newcommand{\sigmabar}{\overline\sigma}
\newcommand{\muhat}{\hat\mu}
\newcommand{\ytilde}{\tilde y}
\newcommand{\wtilde}{\tilde w}
\newcommand{\btilde}{\tilde b}
\newcommand{\Seff}{S_{\rm eff}}
\newcommand{\betatilde}{\tilde\beta}
\newcommand{\ltilde}{\tilde\lambda}
\newcommand{\rtilde}{\tilde r}

\begin{document}

\title{{\bf On the Eichten--Preskill Proposal for Lattice Chiral Gauge
Theories}\thanks{Talk presented by M.~Golterman at the Workshop on
Non-Perturbative Aspects of Chiral Gauge Theories, Rome, March, 1992}}

\author{Maarten~F.~L.~Golterman, Donald~N.~Petcher and Elena~Rivas}
\address{Department of Physics, Washington University, St. Louis, MO 63130,
USA}
\preprint{Wash. U. HEP/92-82}
\date{June, 1992}

\begin{abstract}
We have studied the Eichten--Preskill proposal for
constructing lattice chiral gauge theories using both strong and weak
coupling methods.  The results indicate that this proposal is unlikely
to work due to a dynamical
behavior similar to that of the Smit--Swift proposal, which also does not
give rise to chiral fermions.
\end{abstract}

\maketitle

\section{\bf Introduction}
In a challenging paper \cite{EichPres} a number of years
ago Eichten and Preskill
proposed a method for constructing asymptotically free chiral gauge theories
(AFChGT's) on the lattice.  The idea is to employ Wilson fermions, but
in contrast to the case of vector-like gauge theories,
where a gauge invariant Wilson mass term can be constructed,
the Wilson mechanism is made to work through gauge invariant
four-fermion couplings \cite{Ginsparg}.  They started from a model
containing only left-handed fermion fields, but of course, due to the
species doubling phenomenon \cite{NN,KS}, extra left-handed and
right-handed fermion species appear.  The goal of the four-fermion
interactions
is to dynamically produce fermionic bound states of a handedness
opposite to that of the species doublers and give the ensuing Dirac
particles a mass of the order of the cutoff (i.e. the inverse lattice
spacing), at least in some region of the phase diagram, while the
physical fermion is kept massless.

They showed
that at strong coupling indeed these bound states may be obtained easily
for all fermion species.  The speculation was then that by adjusting the
couplings, one could make the right-handed partner of the physical
fermion disassociate, while at the same time the bound state partners of
the doublers would remain bound. Thus the physical left-handed fermion
would have to be massless and appear in the physical spectrum.
The difference from lattice QCD with Wilson fermions is that in that
case all fermion fields are Dirac fields so that there is no need for
such bound states.

In addition, they assumed it important that the symmetry
structure of the lattice action exactly reproduce that of the target
continuum AFChGT by construction.  This implies that one has
to start from a combination of gauge group and fermion representation
which is anomaly free, and moreover, that any
global symmetry which is anomalous in the corresponding continuum AFChGT
has to be broken explicitly on the lattice.
This was accomplished by a judicious choice of the four-fermion
interaction terms, which
employ all components of the anomaly free fermion
representation in an essential way
and in addition provide all necessary explicit symmetry breaking.

Although this approach is different from that proposed by Swift and
Smit \cite{Swift,SmitZak}
(see for reviews refs. \cite{MGreview,BockJapan}), there are
also similarities, and the question arises whether this proposal can
succeed where the Smit--Swift method failed to produce continuum
chiral gauge theories
\cite{BockJapan,GolPetSmit,BockDeSmit,dnprome}.  In this talk we will
consider an $SO(10)$ chiral gauge theory with left-handed Weyl fermions
transforming in the $16$-dimensional spinor representation,
which of course is free of gauge anomalies.  In their original paper,
Eichten and Preskill proposed an explicit model for an $SU(5)$ AFChGT,
with fermions in the $5$, $\overline{10}$ and singlet representations.  This
model can be obtained from the model of this talk by adding terms
which break $SO(10)$ to $SU(5)$, but, as they already pointed out,
these terms
are irrelevant for the question as to whether the proposal will work or
not.

As an additional simplification we will turn off the gauge fields, since
the proposed mechanism is supposed to work as a consequence of the
four-fermion interactions, which are specifically designed for this
purpose.
  If the gauge fields are
turned off, the target continuum theory describes just undoubled
free Weyl fermions, and the
goal will then be to see whether such a continuum limit can be obtained from
the Eichten--Preskill approach.

A preliminary report on this work appeared in ref. \cite{UsJapan}, for a
more complete report and many details, see ref. \cite{UsEP}.

\section{\bf Review of strong coupling results for the Eichten--Preskill model}
The Eichten--Preskill model for $SO(10)$ with the gauge interaction turned
off is given by the lagrangian
\begin{eqnarray}
\lefteqn{L_{4f} = \half\sum_\mu\left[\psi^\dagger_{Li\;x}\sigmahat_\mu
\psi_{Li\;x+\mu}-\psi^\dagger_{Li\;x+\mu}\sigmahat_\mu\psi_{Li\;x}\right]}
\nonumber \\
 & &\mbox{}-\frac{1}{24}\lambda\left[(\psi^T_{Li}\sigma_2 T^{aij}\psi_{Lj})^2
+\hbox{\rm h.c.}\right] \nonumber \\
 & &\mbox{}-\frac{r}{48}\left[\Delta(\psi^T_L\sigma_2 T\psi_L)^2+\hbox{\rm
h.c.}\right],
\label{eq:4faction}
\end{eqnarray}
where $\sigmahat_\mu=(1, i\sigma_i)$ with $\sigma_i$ the Pauli matrices,
and
$T^{aij}$ an
$SO(10)$ invariant tensor with $a=1,\dots,10$, and
$i,\;j=1,\dots,16$.  In this talk we will set the lattice spacing equal
to one.  $\Delta$ is a symmetric shift
operator leading to Wilson-like nearest neighbor interactions:
\begin{eqnarray}
\lefteqn{\Delta(\psi_i\psi_j\psi_k\psi_l)_x = -{1\over 2}\sum_{\pm\mu}
\lbrack\psi_{i\;x+\mu}\psi_{j\;x}\psi_{k\;x}\psi_{l\;x}} \nonumber \\
 & &\mbox{}+\psi_{i\;x}\psi_{j\;x+\mu}\psi_{k\;x}\psi_{l\;x} \nonumber \\
 & &\mbox{}+\cdots -4\;\psi_{i\;x}\psi_{j\;x}\psi_{k\;x}\psi_{l\;x}\;\rbrack.
\label{eq:delta}
\end{eqnarray}
This model is exactly the model which was considered in ref.
\cite{EichPres}, but without the additional terms which break $SO(10)$
to $SU(5)$.

The four-fermion interactions
explicitly break left-handed fermion number, which is anomalous in
the continuum.  Therefore the problem raised by Banks \cite{Banks,DugMan}
does not arise in this model.

Eichten and Preskill computed the fermion propagator for $r=0$ and strong
$\lambda$ by hopping parameter expansion
techniques \cite{EichPres}, and showed that at
strong coupling a right-handed fermionic bound state exists with the quantum
numbers of $T^{\dagger
a}\psi_L^*(\psi^\dagger_L\sigma_2 T^{\dagger a}\psi_L^*)$.  This implies that
at strong coupling the fermions are Dirac fermions.  The
$LL$ component of the fermion
propagator for the $SO(10)$ model of eq. (\ref{eq:4faction})
with
$\alpha=\frac{1}{\sqrt{\lambda}}$ is (to leading order)
\begin{equation}
S_{LiLj}(p)=\frac{-\frac{1}{\alpha}\sum_\mu \sigmabar_\mu i\sin{p_\mu}
\delta_{ij}}{\sum_\mu\sin^2{p_\mu}+\frac{320}{3\alpha^4}}+O(\alpha^7),
\label{eq:fpropff}
\end{equation}
in which $\sigmabar_\mu=(1,-i\sigma_i)$.
Hence a Dirac fermion is found for strong $\lambda$ with mass
\begin{equation}
am_f=8\sqrt{\frac{5}{3}}\lambda+O\left(\frac{1}{\lambda}\right)
\label{eq:fermionmass}
\end{equation}
($a$ is the lattice
spacing, $m_f$ is the fermion mass in physical units).   Of course, eq.
(\ref{eq:fpropff}) exhibits the species doublers which are all
degenerate at $r=0$.

They then speculated, still with $r=0$,
that a phase transition must exist separating this
strong coupling region from weak coupling perturbation theory where the
fermions are massless and doubled.  At this phase transition, the bound state
would reach threshold, disappear, and leave us with a (doubled) massless
left-handed fermion.  When the Wilson four-fermion coupling $r$ is
turned on, the fermion
spectrum at strong coupling becomes non-degenerate, and
the possibility arises that
this phenomenon would happen only for the lowest lying fermion,
while the doublers would remain heavy.  At or across this phase
transition on the weak coupling side and for $r$ sufficiently strong, thus
a single, massless left-handed Weyl fermion would emerge, with the doublers
still at masses of the order of the cutoff.
An important test of this scenario is to see whether at $r=0$ the phase
transition that must occur between large and small $\lambda$ is
characterized by the (in this case doubled) fermion becoming massless.

A discussion of what might go wrong with these speculations was
also given in ref.~\cite{EichPres}.  At the phase transition
separating strong and weak
coupling the $SO(10)$ symmetry could be spontaneously broken, or {\it
all}
fermions including the doublers might become massless
simultaneously at this transition
despite their  non-degenerate spectrum
 at strong coupling.

In this context it is interesting to note that
both of these unwanted
phenomena actually happen in the Smit--Swift model
\cite{BockJapan,GolPetSmit,Aachenphase} which was designed to
yield chiral fermions in a comparable way through strong Wilson--Yukawa
couplings to scalar fields.
In that model, a ferromagnetic (FM)
or broken symmetry phase
separates different paramagnetic or  symmetric phases at strong and weak
coupling (PMS and PMW respectively).
The PMS phase occurs for large values of the
Yukawa couplings (at least one of the two Yukawa couplings, the
single-site Yukawa coupling or the Wilson--Yukawa coupling has to be
large).  Fermions are typically massive, with masses of the order of the
cutoff, in the PMS phase.
At the PMS-FM transition the symmetry to be gauged is
broken spontaneously while the fermions remain massive.
Then at
the FM-PMW transition all fermions become massless simultaneously, and hence
no region with chiral fermions and without doublers exists.
(At the FM-PMW transition the gauge symmetry is restored.)
The fact that for
strong Wilson--Yukawa coupling one fermion can be tuned to be massless
\cite{GolPetshift} does not affect this conclusion \cite{GolPetSmit}.
For a review of the dynamics of the Smit--Swift model, see
ref. \cite{dnprome} in these proceedings.

In fact, a result was already presented
in ref.~\cite{EichPres} which indicates that
for $r=0$ this is precisely what seems to happen in the model at hand.
A scalar bound state with
interpolating field
\begin{equation}
A_1^a =\psi_L^T\sigma_2 T^a\psi_L+\hbox{\rm h.c.}
\label{eq:Adef}
\end{equation}
exists at strong $\lambda$ and the
propagator  $G$ for this scalar field to leading order is
\begin{equation}
G_{ab}(p)=\frac{\frac{256}{\alpha^2}\delta_{ab}}{4\sum_\mu\sin^2{\half p_\mu}
+\frac{40}{3\alpha^2}-8}.
\label{eq:bpropff}
\end{equation}
{}From this equation the boson mass is
\begin{equation}
(am_b)^2=8\lambda\left(\frac{5}{3}-\frac{1}{\lambda}\right)
\label{eq:bosonmass}
\end{equation}
to leading order,
which becomes negative for small enough $\lambda$
(to this order $\lambda<\lambda_c=0.6$), indicating
spontaneous symmetry breaking.  As this scalar field transforms in the
ten dimensional defining representation of $SO(10)$, one expects the
symmetry to be broken to $SO(9)$.  Note that,
from eq. (\ref{eq:fpropff}), consistent to this order, the
fermion stays massive at this transition.  Since the fermion is massive
on the symmetric side of the phase transition (i.e. in the PMS phase),
there is no reason for the fermion mass to vanish at the phase
transition.

It turns out that another scalar field appears in the model defined by eq.
(\ref{eq:4faction}), which also transforms in the ten dimensional
representation, but has different quantum numbers under $CP$.  This
scalar field turns out to produce negative metric states, which to
leading order are degenerate in mass with the field $A_1$.  If this
would persist to all orders, the model would have to be modified in
order to cure this disease.  For details, see ref. \cite{UsEP}.

\section{\bf An equivalent Higgs--Yukawa model}
We have studied the nature of the strong coupling phase transition in a
closely
related model, in which the four-fermion couplings have been replaced by Yukawa
couplings and hence a scalar field $\phi_1$ with the same quantum
numbers as the field $A_1$ defined in eq. (\ref{eq:Adef}) appears
explicitly:\footnote{A field for the other scalar excitation mentioned at the
end
of the previous section should in fact be included, but we will omit
that here, see ref. \cite{UsEP}}
\begin{eqnarray}
\lefteqn{L_{Y} = \half\sum_\mu\left[\psi^\dagger_{Li\;x}\sigmahat_\mu
\psi_{Li\;x+\mu}-\psi^\dagger_{Li\;x+\mu}\sigmahat_\mu\psi_{Li\;x}\right]}
\nonumber \\
 & &\mbox{}+\half
b\phi^a_{1\;x}\phi^a_{1\;x}-\kappa\sum_{\mu}\phi^a_{1\;x}\phi^a_{1\;x+\mu}
\nonumber \\
 & &\mbox{}+(\half y\psi^T_L\sigma_2 T^a\psi_L\phi_1^a+\hbox{\rm h.c.})
\nonumber \\
 & &\mbox{}-(\frac{1}{4} w\psi^T_L\sigma_2 T^a\Box\psi_L\phi_1^a+\hbox{\rm
h.c.}),
\label{eq:wyaction}
\end{eqnarray}
where $\Box$ is the lattice laplacian and $\phi_1$ is a real scalar field.
Sums over repeated indices $a$ are implied.
We have not explicitly indicated a possible additional scalar potential.
The Yukawa couplings $y$ and
$w$ again break the fermion number symmetry.
For $\kappa=0$ and $w=0$ this model is just that of eq.
(\ref{eq:4faction}) for $r=0$ and we believe that also for $\kappa$ and $w$
non-vanishing this model is in the same universality class as
the model of eq. (\ref{eq:4faction}) \cite{UsEP}.

We have studied this model at strong coupling by employing
a large $N$ limit \cite{Kuti} suited for large values of $y$ and $w$.
The fermion fields are given an additional index running from $1$ to
$N$, which introduces an extra (anomaly free) $SO(N)$ symmetry into the
model.  Rescaling the couplings with $N$
\begin{equation}
y=\ytilde\sqrt{N},\ \ \ \ \ w=\wtilde\sqrt{N},\ \ \ \ \ b=\btilde N,
\label{eq:rescaling}
\end{equation}
keeping $\ytilde$, $\wtilde$, $\btilde$ and $\kappa$ fixed, the
effective action $\Seff$
for the scalar field $\phi_1$ can be computed to
leading order in $1/N$, with the result
\begin{equation}
\Seff=-(\kappa+\frac{1}{8}\btilde\betatilde^2 J(\betatilde\wtilde))
\sum_{x,\mu}\phi^a_{1\;x}\phi^a_{1\;x+\mu},
\label{eq:seff}
\end{equation}
with constraint
\begin{equation}
\phi^a_{1\;x}\phi^a_{1\;x}=1.
\label{eq:constraint}
\end{equation}
In the equation for $\Seff$,
\begin{equation}
\betatilde=\frac{1}{4\wtilde+\ytilde}
\label{eq:beta}
\end{equation}
and
\begin{equation}
J(\betatilde\wtilde)=\int_p\frac{\sum_\mu\sin^2{p_\mu}}
{\left(1-\betatilde\wtilde\sum_\mu
\cos{p_\mu}\right)^2},
\label{eq:int}
\end{equation}
where
\begin{equation}
\int_p\equiv\frac{1}{(2\pi)^4}\int\int\int\int_{-\pi}^{\pi}d^4p.
\label{eq:intdef}
\end{equation}
This implies that, for $\kappa\le\kappa_c$, there will be a second order phase
transition line in the $y-w$ plane separating a PMS phase from an FM phase.
The location to leading order in $1/N$ is given by
\begin{equation}
8\btilde\betatilde^2 J(\betatilde\wtilde)=\kappa_c-\kappa,
\label{eq:phasetrans}
\end{equation}
with $\kappa_c$ the critical value for $\kappa$ in the purely bosonic $SO(10)$
spin model.

We may also derive an expression for the fermion propagator at strong
$y$ and/or $w$ (corresponding to strong $\lambda$ and/or $r$) for the
Higgs--Yukawa model.  For instance, for the $LL$ component we find,
ignoring corrections of order $1/N^2$
\begin{equation}
S_{LL}(p) =
{{-i\sum_\mu\sigmabar_\mu\sin{p_\mu}}\over {M^2(p)+z^2
\sum_\mu\sin^2{p_\mu}}},
\label{eq:fermionprop}
\end{equation}
where
\begin{equation}
M(p)=y+w\sum_\mu(1-\cos{p_\mu}),
\label{eq:Mdef}
\end{equation}
and
\begin{equation}
z^2={{\tilde
b}\over{32}}\langle\phi^a_1(x)\phi^a_1(x\pm\muhat)\rangle.
\label{eq:zsqdef}
\end{equation}
Since $z^2$ is finite and non-vanishing at the phase transition eq.
(\ref{eq:phasetrans}),
we find, as in the case of the Smit--Swift model,  one massless Dirac
fermion and doublers with masses of the order of the cutoff (for
$y\to 0, w$ finite).  However, none of the bound state partners of the fermion
species disassociate at the phase transition, and all fermions in the
emerging spectrum are of a Dirac-like nature.  (The composite operator
for the right-handed fermion field is $\phi_1^a(x)T^{\dagger
a}\sigma_2\psi^*(x)$ for the Higgs--Yukawa version of the model.)

Note that the large $N$ technique used here is very similar to the
$1/w$ expansion developed in ref. \cite{Usoneoverw}.  As in that case, the
approach allows us to integrate out the fermions systematically through
perturbation theory in a small parameter (here $1/N$), but it does not
provide us with a way of computing the bosonic correlation functions
that remain, as for instance eq. (\ref{eq:zsqdef}).
To this order in $1/N$, this
is however the only bosonic correlation function we need.

\section{\bf Results at weak coupling}
In this section we consider the weak coupling side of the phase diagram,
employing techniques similar to those used in ref. \cite{Kuti}.  At weak
coupling, the doublers will of course always be present in the physical
spectrum, but nevertheless we can learn two useful things.  First, we
find that for increasing values of $\lambda$ and $r$ (or $y$ and $w$ in
the Higgs--Yukawa version) spontaneous symmetry breaking occurs, giving
independent confirmation that indeed a broken phase separates the two
symmetric phases at strong and weak coupling.  Second, we can check that
both models are in the same universality class.  Again, details can be
found in ref. \cite{UsEP}. Here we only present the main results.

At weak coupling both the four-fermion and the Higgs--Yukawa versions of
the model can be solved exactly in large $N$, if one defines
\begin{equation}
\ltilde=\lambda N,\ \ \ \ \ \rtilde=rN
\label{eq:ffcouplings}
\end{equation}
in the case of the four-fermion model, and
\begin{equation}
\ytilde=y\sqrt{N},\ \ \ \ \ \wtilde=w\sqrt{N}
\label{eq:hycouplings}
\end{equation}
in the case of the Higgs--Yukawa model.

In the four-fermion model, defined through the lagrangian in
eq. (\ref{eq:4faction}), the result for the
inverse fermion propagator is
\begin{eqnarray}
\lefteqn{S^{-1}(p)=   }\nonumber \\
\left(\begin{array}{cc}
i\sum_\mu \sigmahat_\mu \sin p_\mu & \sum_a\eta^a T^\dagger_a
\Sigma(p)\\
\sum_a\eta^a T_a \Sigma(p)&i\sum_\mu \sigmabar_\mu \sin p_\mu
\end{array}\right),
\label{eq:invprop}
\end{eqnarray}
where $\Sigma$ satisfies the gap equation
\begin{eqnarray}
\lefteqn{\Sigma(p)=  }\nonumber \\
 & &\frac{16}{3}\int_q\frac{\Sigma(q)}{D(q)}[\ltilde+
\rtilde\sum_\mu(2-\cos{p_\mu}-\cos{q_\mu})],
\label{eq:gap}
\end{eqnarray}
with
\begin{equation}
D(q)\equiv\sum_\mu\sin^2{q_\mu}+\Sigma^2(q).
\label{eq:D}
\end{equation}
The vector $\eta^a$ indicates the
direction of (possible) symmetry breaking in the ten dimensional space.

For small values of $\ltilde$ and $\rtilde$, we are in the symmetric
phase and $\Sigma(p)=0$. There is second order phase transition to a
spontaneously broken symmetry phase with $\Sigma(p)\ne 0$ at
\begin{equation}
\ltilde=\frac{1-8\rtilde I-(4I^2-\frac{16}{3}I)\rtilde^2}{I},
\label{eq:weakphtr}
\end{equation}
where
\begin{equation}
I=\frac{16}{3}\int_q\frac{1}{\sum_\mu\sin^2{q_\mu}}=3.31\;.
\label{eq:I}
\end{equation}
At $r=0$ this leads to a critical value for $\lambda$ of $0.3/N$ for
the PMW-FM phase transition, which even at $N=1$ is much smaller than
the strong coupling value of $0.6$ for the FM-PMS transition.  (The
large $N$ technique used at strong coupling does not apply to the
four-fermion version.)

The propagator for the boson field $\phi_1$ may then be computed, with
the result that across the phase transition nine Goldstone bosons
emerge, as expected.  Also, in the broken phase one finds that the
renormalized Yukawa coupling is given by the ratio of the renormalized
fermion mass and scalar vacuum expectation values.

These calculations can be repeated in the Higgs--Yukawa version of the
model, with identical results.  In fact, the correlation functions can,
in the scaling region around the phase transition line, be mapped onto
each other if one identifies
\begin{eqnarray}
\frac{\rtilde}{\rho}&=&\frac{\wtilde}{\ytilde},\nonumber \\
\kappa&=&0,
\label{eq:matching}
\end{eqnarray}
where
\begin{equation}
\rho^{-1}=\frac{16}{3\Sigma(0)}\int_q\frac{\Sigma(q)}{D(q)}
\label{eq:rho}
\end{equation}
is a function of $\ltilde$ and $\rtilde$.
This confirms our belief that the
four-fermion and Higgs--Yukawa models are indeed in the same
universality class.

\section{\bf Conclusion}
Although our results here are mostly obtained from various large $N$
computations, we believe that they represent good evidence that the phase
diagram of the $SO(10)$ version of the Eichten--Preskill proposal looks very
similar to the phase diagram of Smit--Swift type models.  In particular, rather
than producing a region where undoubled left-handed Weyl fermions exist, the
$SO(10)$ symmetry breaks spontaneously, while the fermions, which acquire
masses of the order of
the cutoff in the strong coupling region, remain massive.

At strong coupling there exists the possibility of tuning the physical
fermion to be massless while the doublers remain heavy, but the bound
state partners of all the fermion species remain bound across the
symmetry breaking phase transition.  They become unbound at a second
phase transition at much weaker couplings, where the symmetry is
restored, and at that phase transition all fermion species become
massless simultaneously, apparently leaving no room for an undoubled,
chiral fermion spectrum.

When the $SO(10)$ gauge fields are turned on, the emerging gauge theory
will have vectorlike instead of chiral couplings.  This is because
all fermions transform in $SO(10)$ spinor
representations, including the right-handed fermions which exist at the
PMS-FM transition.  The situation is similar to that in the ``modified"
Smit--Swift model \cite{GolPetSmit} (we have checked this explicitly at
strong coupling).

The Eichten--Preskill proposal is a proposal starting from a very
careful analysis of the anomaly structure of the target AFChGT
(see also the contribution of Banks to these proceedings), unlike
the Smit--Swift approach (but see also refs. \cite{DugMan,JanSeillac}).
Whatever one may think of this, it is clear that if the
Eichten--Preskill model fails to yield chiral fermions because of the
arguments presented here, these reasons have little to do with anomalies.
No triangle loops ever appeared in our computations, neither in the
analysis of the Smit--Swift model, nor in that of the Eichten--Preskill
model.

Of course, our analytic results do not fully cover the
phase diagram, and our conclusions rest on extrapolation of these results.
In principle, the middle region of the phase diagram could turn out to be
complicated, and we cannot logically exclude that chiral fermions would emerge.
However, this would then have to occur for reasons beyond the
scenario presented in ref. \cite{EichPres}.  Experience with other
Higgs--Yukawa models such as the Smit--Swift model indicates that the
phase diagram may indeed be more complicated than what can be learned
from our analytic techniques, but
nonetheless, at the same time this more complicated
structure has not led to the emergence of chiral fermions.

\begin{acknowledge}
We thank the organizers of the ``Rome workshop" for an extremely
interesting and pleasant workshop.
We also thank the Aspen Center for Physics, where
part of this research was done.
M.G. and D.P. are partially supported by the US Department of Energy,
and E.R. is supported by a Formaci\'on del Personal Investigador
fellowship from the
Spanish government.
\end{acknowledge}


\end{document}